\documentclass[10pt,letterpaper,twocolumn]{article} 

\usepackage{ol2}
\usepackage[draft]{hyperref}
\usepackage{amsmath}
\usepackage{amssymb}
\begin{document}

\twocolumn[ 

\title{Improved High Contrast Imaging with On-Axis Telescopes\\
using a Multi-Stage Vortex Coronagraph}


\author{Dimitri Mawet,$^{1,*}$ Eugene Serabyn,$^1$ J. Kent Wallace, $^1$ Laurent Pueyo$^{1,2,3}$}

\address{
$^1$Jet Propulsion Laboratory, California Institute of Technology,
4800 Oak Grove Dr, Pasadena, CA 91109, USA \\
$^2$Space Telescope Science Institute, 3700 San Martin Drive, Baltimore, MD 21218, USA\\
$^3$Johns Hopkins University, Physics \& Astronomy, 366 Bloomberg Center,\\
3400 N. Charles Street, Baltimore, MD 21218, USA\\
$^*$Corresponding author: dmawet@eso.org 
}

\begin{abstract}The vortex coronagraph is one of the most promising coronagraphs for high contrast imaging because of its simplicity, small inner working angle, high throughput, and clear off-axis discovery space. However, as with most coronagraphs, centrally-obscured on-axis telescopes degrade contrast. Based on the remarkable ability of vortex coronagraphs to move light between the interior and exterior of pupils, we propose a method, based on multiple vortices, that, without sacrificing throughput, reduces the residual light leakage to $(a/A)^n$, with $n \ge 4$, and $a$ and $A$ being the radii of the central obscuration and primary mirror, respectively. This method thus enables high contrasts to be reached even with an on-axis telescope.\end{abstract}


  ]

\noindent Coronagraphy enhances contrast in astrophysical scenes where the goal is imaging faint features near bright objects. The typical example is an extrasolar planet located less than an arcsecond from its much brighter host star.  The vortex coronagraph, or VC \cite{Mawet2005,Foo2005,Jenkins2008,Swartzlander2009}, can observe very close to bright stars, and so is one of the most promising of coronagraph types. The vector vortex coronagraph, or VVC, which is a realization of the vortex coronagraph based on a circularly symmetric halfwave plate \cite{Mawet2009a}, has been extensively demonstrated in the lab, in visible \cite{Mawet2009b} and near-infrared \cite{Mawet2010} light. Two VVC masks have been installed on the 200-inch Hale telescope at Palomar, and have been used behind an off-axis unobscured subaperture corrected with adaptive optics to successfully image brown dwarfs and extrasolar planets \cite{Mawet2010, Serabyn2010}. However, as with any coronagraph, additional diffraction due to the secondary mirror obstruction present in on-axis telescopes will degrade achievable contrast. Here we show that this extra leakage can be greatly reduced by using two vortex coronagraph stages in sequence.

\begin{figure}[!b]
\centerline{\includegraphics[width=8cm]{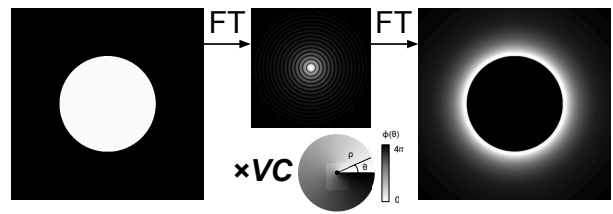}}
\caption{Illustration of the diffraction effect of the vortex phase mask on a filled aperture (left). All of the on-axis coherent light appears outside of the geometric image of the input pupil (right). A circular aperture (Lyot stop) then blocks it all. FT stands for Fourier transform.\label{fig1}}
\end{figure}

A vortex coronagraph applies a helical phase of the form $e^{i\phi}$, with $\phi=l\theta$, where $\theta$ is the focal plane azimuthal coordinate and $l$ the (even) vortex ``topological charge'', to a telescope's focal-plane field distribution, by means of a transparent phase mask (Fig.~\ref{fig1}). The light then propagates downstream to a pupil image (the ``Lyot'' plane), where for an ideal unobscured circular input pupil (Fig.~\ref{fig1}, left), all of the (on-axis) starlight appears outside of the geometric image of the input pupil, rendering the interior of the output pupil empty of the incoming on-axis illumination (Fig.~\ref{fig1}, right). A Lyot-plane aperture then blocks all of the diffracted starlight. Light from off-axis objects misses the center of the vortex and propagates normally.
 
The effect of a charge $l=2$ vortex phase, $e^{i2\theta}$, applied to the ideal focal plane field (Airy pattern), $\frac{2J_1(k\rho A)}{k\rho A}$, of a filled circular aperture of radius $A$, where k is the wavenumber and $\rho$ is the radial coordinate in the focal plane,  has been calculated analytically \cite{Mawet2005, Jenkins2008}, and the Fourier transform of $e^{i2\theta}\times\frac{2J_1(k\rho A)}{k\rho A}$ to the pupil (Lyot) plane yields
\begin{equation}\label{pup}
E_{Lyot}(r,\psi)=\left \lbrace
\begin{array}{ll}
0 & r<A \\
e^{i2\psi}\left(\frac{A}{r}\right)^2 & r>A
\end{array}
\right.
\end{equation}

\noindent 
The field outside the pupil area thus has a $1/r^2$ falloff in this case. 
\begin{figure*}[!t]
\centerline{\includegraphics[width=14.0cm]{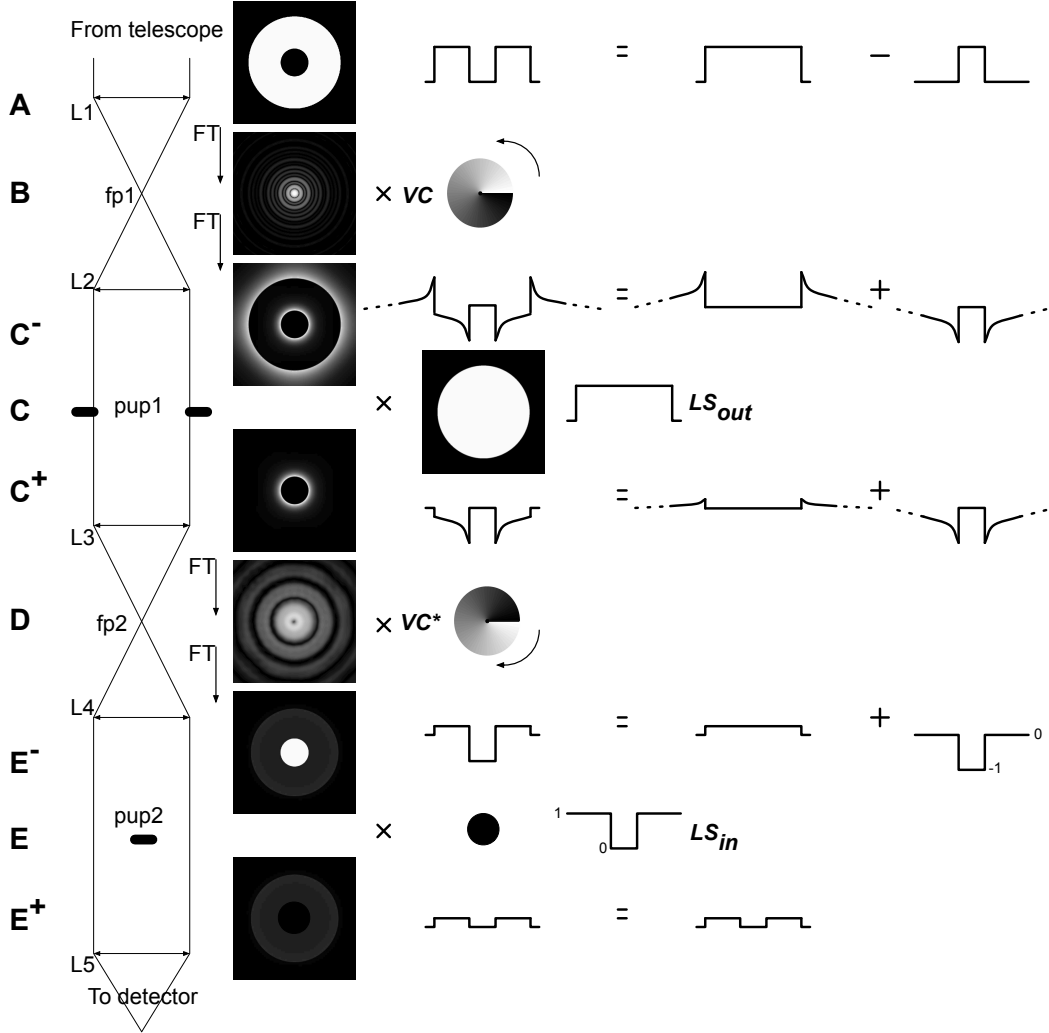}}
\caption{Layout of the two-stage vortex coronagraph. \textbf{A}. Obscured entrance pupil decomposed into the difference of
two filled apertures with radii $A$ and $a<A$. \textbf{B}. Application of the Vortex Coronagraph ($VC$) phase ramp $e^{i\phi}$ to the image produced by the imaging lens L1. \textbf{C$^-$}. The pupil plane distribution after pupil imaging lens L2. The crosscuts show the linear decomposition of the field into the sum of a positive term $(A/r)^2$ for the filled pupil of radius $A$, and another negative term $-(a/r)^2$ corresponding to the filled pupil of radius $a<A$. \textbf{C}. First Lyot stop $LS_{out}$. \textbf{C$^+$}. Post Lyot stop field. The crosscuts show the decomposition of the residual field as the sum of the negative pupil term $-(a/r)^2$, and a second term that leads to cancellation beyond $A$. \textbf{D}. Image in the second focal plane after L3; note that the broad distribution now arises predominantly from the effect of $a$. Also shown is the conjugated vortex $VC^*$, with a phase ramp $e^{-i\phi}$. \textbf{E$^-$}. Pupil plane field after the second vortex, decomposed into the sum of an outer pupil of radius $A$ with a reduced amplitude $(a/A)^2$, and the inner pupil of radius $a$ of amplitude $-1$. \textbf{E}. Application of the second, inner Lyot stop ($LS_{in}$). \textbf{E$^+$}. The final exit pupil after the second stage is a replica of the original entrance pupil left with an intensity reduced by the factor $(a/A)^4$. \label{fig2}}
\end{figure*}

Including now a central obstruction of radius $a$, an on-axis telescope has a pupil field distribution given by
\begin{equation}\label{pup_cent}
E_{pup}(r,\psi)=\left \lbrace
\begin{array}{ll}
0 & r<a \\
1 & a < r<A \\
0 & r > A
\end{array}
\right.
\end{equation}

\noindent
Using the superposition principle, a centrally obscured pupil can be seen as the difference between a filled pupil of radius $A$ and a smaller filled pupil of radius $a$ (Fig.~\ref{fig2},  \textbf{A}), with the field distribution 
\begin{equation}\label{pup_centbis}
E_{pup}(r,\psi)=\left \lbrace
\begin{array}{ll}
-1 & r<a \\
0 & r>a 
\end{array}
\right \rbrace + \left \lbrace
\begin{array}{ll}
1 & r<A \\
0 & r>A
\end{array}
\right \rbrace
\end{equation}

\noindent
Applying the same process to each component yields
\begin{equation}\label{pup_cent_lyot1}
E_{pup1}(r,\psi)=\left \lbrace
\begin{array}{ll}
0 & r< a \\
-e^{i2\psi}\left(\frac{a}{r}\right)^2 & a<r<A \\
e^{i2\psi}\left[\left(\frac{A}{r}\right)^2-\left(\frac{a}{r}\right)^2\right] & r>A
\end{array}
\right.
\end{equation}

\noindent
The field beyond radius $A$ can be removed by the opaque "Lyot" stop (Fig.~\ref{fig2}, \textbf{C}), leaving the post Lyot stop field
\begin{equation}\label{pup_cent_lyot2}
E_{pup1}(r,\psi)=\left \lbrace
\begin{array}{ll}
0 & r< a \\
-e^{i2\psi}\left(\frac{a}{r}\right)^2 & a<r<A \\
0 & r > A
\end{array}
\right.
\end{equation}

\noindent
The residual field interior to the pupil (between $a$ and $A$) leads to contrast degradation in the subsequent focal plane image, as the fraction of the total energy remaining inside the pupil is $(a/A)^2$, or $1\%$ for a $10\%$ central obscuration. 

We now show that a pair of conjugated vortex coronagraphs in series reduces the light leakage due to the central obscuration by an additional factor of $(a/A)^2$, leading to a total  leakage of $(a/A)^4$. Our approach is based on the remarkable ability of the vortex mask to move the light between the interior and exterior of the pupil (Eqn. \ref{pup}). Using superposition and the self-similarity of the $1/r^2$ terms, the field after the first Lyot stop can be rewritten (Fig.~\ref{fig2}, \textbf{C}$^+$)
\begin{equation}\label{pup_cent_lyot2bis}
E_{pup1}(r,\psi)=\left \lbrace
\begin{array}{ll}
0 & r< a \\
-e^{i2\psi}\left(\frac{a}{r}\right)^2 & a<r<A \\
e^{i2\psi}\left(\frac{a}{A}\right)^2\left(\frac{A}{r}\right)^2 -e^{i2\psi}\left(\frac{a}{r}\right)^2  & r>A \\
\end{array}
\right.
\end{equation}

We now go to a second stage focal plane by Fourier transforming Eqn. \ref{pup_cent_lyot2bis}, yielding 
\begin{equation}\label{im2}
E_{fp2}(\rho,\theta)=-e^{i2\theta}\frac{2J_1(k\rho a)}{k\rho a}+ e^{i2\theta}\frac{2J_1(k\rho A)}{k \rho A} \left(\frac{a}{A}\right)^2
\end{equation}

\noindent
Multiplying this function by the ${\it conjugated}$\footnote{The VV mask applies opposite phase ramps $e^{i\phi}$ and $e^{-i\phi}$ to the orthogonal circular polarization states and also interchanges the polarization states \cite{Mawet2005,Mawet2009a}. Thus a second VV mask downstream from the first applies the conjugated phase ramp automatically.} vortex $e^{-i2\theta}$ cancels the phase term (Fig.~\ref{fig2}, \textbf{D}), leaving
\begin{equation}\label{im2}
E_{fp2}(\rho,\theta)=-\frac{2J_1(k\rho a)}{k\rho a}+ \frac{2J_1(k\rho A)}{k \rho A} \left(\frac{a}{A}\right)^2
\end{equation}

\noindent
Transforming back to a second pupil plane we get (Fig.~\ref{fig2}, \textbf{E}$^-$)
\begin{equation}\label{pup_cent_lyot3}
E_{pup2}(r,\psi)=\left \lbrace
\begin{array}{ll}
-1 & r<a \\
0 & r>a
\end{array}
\right \rbrace + \left \lbrace
\begin{array}{ll}
\left(\frac{a}{A}\right)^2 &r<A \\
0 & r > A
\end{array}
\right \rbrace
\end{equation}

\noindent
The bright central region is then blocked by a second Lyot stop (Fig.~\ref{fig2}, \textbf{E}), of radius $a$, which leaves 
\begin{equation}\label{pup_cent_lyotfin}
E_{pup2}(r,\psi)=\left \lbrace
\begin{array}{ll}
0 & r<a \\
\left(\frac{a}{A}\right)^2 & a < r<A \\
0 & r > A
\end{array}
\right.
\end{equation}

\noindent
The residual field distribution is thus an exact replica of the original distribution (Eqn. \ref{pup_cent}), but with the field reduced by $\left(\frac{a}{A}\right)^2$, and the intensity (per area or integrated) reduced by $\left(\frac{a}{A}\right)^4$ (Fig.~\ref{fig2}, \textbf{E}$^+$). For a $10\%$ secondary, the total residual light is then only $10^{-4}$, and the theoretical focal plane contrast at e.g.~the 3rd Airy ring is $\sim 10^{-7}$. Thus, even with an on-axis telescope, high contrasts are achievable with such a "two-stage vortex" coronagraph. Moreover, by extension, further vortex stages likewise decrease leakage by factors of $(a/A)^2$, leading to a total leakage of $(a/A)^{2n}$, where $n$ is the number of stages\footnote{Note that multi-stage coronagraphy had also been proposed previously as a way to increase the working bandwidth and reduce the sensitivity to tip-tilt and stellar size leakage \cite{Yaitskova2005,Baudoz2008,Baudoz2010,Abe2008}.}, at the cost of additional optics and potential errors (wavefront imperfections, alignment issues, ...). 

Of course secondary mirrors come with support structures (spiders). Examination of the influence of spiders on contrast is beyond the scope of this letter, but initial calculations show that the edge enhancement property of coronagraphs and particularly optical vortices \cite{Swartzlander2009} allows most of the diffracted light to stay within the spiders' physical width, which can then be blocked by a Lyot stop designed to cover them at each stage. As a result, the impact of reasonably sized spiders (e.g. width $\lesssim 1\%$ of the telescope diameter) is below the limitations imposed by the central obstruction. Thus, using a multi-stage vortex, high contrast can be obtained even with on-axis telescopes, important in both the space-based and ground-based cases.

\bigskip

This work was carried out at the Jet Propulsion Laboratory, California Institute of Technology, under contract with NASA. LP acknowledge support from the Carl Sagan Fellowship Program.


\end{document}